\documentclass[twocolumn,showpacs,showkeys]{revtex4}
\usepackage{graphicx}
\usepackage{bm}
\usepackage{color}
\usepackage{amsmath}
\usepackage{natbib}

\begin{document}

\title{Hydrodynamic model of BEC with anisotropic short range interaction and the bright solitons in the repulsive BEC}

\author{Pavel A. Andreev}
\email{andreevpa@physics.msu.ru}
\affiliation{Faculty of physics, Lomonosov Moscow State University, Moscow, Russian Federation.}

\date{\today}

\begin{abstract}
The quantum hydrodynamic model is developed for the axial symmetric anisotropic short-range interaction.
The quantum stress tensor presents the interaction.
It is derived up to the third order by the interaction radius.
The first order by the interaction radius contains the isotropic part only.
It leads to the interaction in the Gross-Pitaevskii approximation.
Terms existing in the third order by the interaction radius are caused by the isotropic and nonisotropic parts of the interaction.
Each of them introduces the interaction constant.
Therefore, three interaction constants are involved in the model.
Atoms, except the alkali and alkali-earth atoms, can have anisotropic potential of interaction,
particularly it is demonstrated for the lanthanides.
The short-wavelength instability caused by the nonlocal terms appears in the Bogoliubov spectrum.
Conditions for the stable and unstable behaviour are described.
Bright solitons in the repulsive BEC are studied under influence of the anisotropic short-range interaction in the BEC of one species.
Area of existence of this bright solitons corresponds to the area of the instability of the Bogoliubov spectrum.
Approximate reduction of the nonlocal nonlinearity to the quintic nonlinearity at the description of the bright solitons is demonstrated either.

\end{abstract}

\pacs{03.75.Hh, 03.75.Kk, 67.85.Pq}
\keywords{bright soliton, repulsive BEC, anisotropic interaction, nonlocal interaction, quantum stress tensor.}


\maketitle


\section{Introduction}

Solitons are famous nonlinear phenomena existing in various mediums.
Being discovered on the water surface,
solitons are found in plasmas \cite{Tran PS 79}, \cite{Lonngren PP iop 83}, \cite{KUZNETSOV PR 86},
in lasers \cite{Garbin Chaos 17},
in ultracold bosonic atoms \cite{Shaukat PRA 18}, \cite{Astrakharchik PRA 18},
in ultracold fermions \cite{Syrwid PRA 18}, \cite{Yefsah nat 13},
in different setups of the condensed matter physics
like spin waves \cite{Boardman PRB 93}, \cite{Bagada PRL 97}, \cite{Wang PRL 11},
and
charge density waves \cite{Rojo-Bravo PRB 16}, \cite{Matsuura EPL 15}, \cite{Brazovskiiand JETP 91}, \cite{Latyshev PRL 05}.
Weak solitons observed over astronomical distances recirculating in an optical fibre loop \cite{Jang NP 13}.
Solitons show different features like attraction, repulsion, breaking up, merging, orbiting one another
and the annihilation of two solitons \cite{Gordon Op L 83}, \cite{Reynaud EPL 90}, \cite{Tikhonenko PRL 96}, \cite{Shih PRL 97}, \cite{Krolikowski Op L 98}. Recently, it is demonstrated that
solitons can absorb the resonant waves \cite{Yulin PRA 18}.

At the analysis of the nonlinear waves,
many physical models of different physical systems
lead to the nonlinear Schrodinger equation (NLSE)
(such as the Gross-Pitasevskii (GP) equation) or the KortewegЦ-de Vries (KdV) equation with cubic nonlinearity,
where the nonlinearity is mostly caused by the interparticle interaction.
For instance, the attractive short-range interaction in Bose-Einstein condensate (BEC) of neutral particles leads
to the bright soliton existence (the area of the concentration amplification)
while the repulsive interaction leads to the dark soliton formation (the area of the concentration rarefication).
It seems that the Fermi pressure is the cause of the dark solitons appearance in the ultracold fermions \cite{Yefsah nat 13},
where the Fermi pressure gives the effective repulsion,
but the nonlinearity is fractional.
However, more generale models contain complex forms of nonlinearities, such as the quintic nonlinearity or the nonlocal nonlinearities.
Generalised models predict existence of the bright soliton in mediums
with repulsive meanfield interaction \cite{Andreev MPL B 12}, \cite{Wang NJP 14}.

There are two kinds of the nonlocal nonlinearities:
the integral nonlinear term, like the dipolar BEC model \cite{Lahaye RPP 09}, \cite{Santos PRL 00}, \cite{Baranov PR 08},
and the nonintegral nonlinearities containing spatial derivatives of the macroscopic wave function \cite{Andreev PRA08}, \cite{Rosanov}, \cite{Braaten}.

A model with nonlocal nonlinearity containing spatial derivatives is suggested
in this paper at the analysis of the anisotropic short range interaction in the atomic BEC.
Particularly, the paper is focused on study of the bright solitons in the repulsive BEC
and the anisotropy of their properties.

Recent experiments with the BEC of rare-earth elements demonstrated phenomena requiring new physics for their explanation.
It seems that it requires consideration of the quantum fluctuations in BEC
in addition to the standard (meanfield) short-range interaction (SRI) and dipole-dipole interaction.
To some extend, the "constructive" three-particle interaction \cite{Kohler PRL 02}, \cite{Braaten PRL 02}, \cite{Bulgac PRL 02}, \cite{Buchler NP 07}, \cite{Petrov PRL 14}, \cite{Andreev IJMPB 13}
described by the fifth-order nonlinearity in the Gross-Pitaevskii equation was applied to these phenomena \cite{Blakie pra 16}.
However, the quantum fluctuations demonstrated better agreement with experimental data.
All mentioned interaction and related to them quantum fluctuations are applicable to the alkali-earth (see Ref. \cite{Lahaye Nat 07} for $^{52}$Cr) atoms and rare-earth atoms (see Ref. \cite{Ferrier-Barbut PRL 16}, \cite{Pfau Nature 16} for dysprosium $^{164}$Dy).
However, the quantum Rosensweig instability is found for the rare-earth atoms.
Hence, the following question can be formulated:
is it quantitative or qualitative difference.
The rare-earth atoms show a qualitative difference due to the anisotropy of the SRI.
It shows that a model of the rare-earth atoms with the account of the anisotropic SRI is required.
The analysis of the spectrum of Feshbach resonances in the rare-earth atoms demonstrates an extra feature:
anisotropy of the quasi-potentials \cite{Petrov PRL 12}.

It is found that the quantum Rosensweig instability in BEC of lanthanides
(which are atoms with relatively large dipole-dipole interaction)
can be modeled by an effective NLSE with the local form of the quantum correlations  \cite{Baillie 16}, \cite{Wachtler 1605},
\cite{Bisset PRA 16}
\begin{equation}\label{GP with qf} \imath\hbar\partial_{t}\Psi=\biggl[-\frac{\hbar^{2}\nabla^{2}}{2m}+g\mid\Psi\mid^{2} +\Phi(\textbf{r},t)+\gamma_{QF}\mid\Psi\mid^{3}\biggr]\Psi, \end{equation}
where
$\Psi$ is the macroscopic (effective) wave function,
\begin{equation}\label{} \Phi(\textbf{r},t)= \mu^{2}\int d\textbf{r}'
\frac{1-3\cos^{2}\theta'}{|\textbf{r}-\textbf{r}'|^{3}}\mid\Psi(\textbf{r}',t)\mid^{2} \end{equation}
is the potential of the magnetic dipole-dipole interaction \cite{Lahaye RPP 09}, \cite{Santos PRL 00}, \cite{Baranov PR 08}, \cite{Wachtler 1601},
\begin{equation}\label{gamma QF}
\gamma_{QF}=\frac{32}{3}g\frac{a_{s}^{3}}{\pi}\biggl(1+\frac{3}{2}\epsilon_{dd}^{2}\biggr)\end{equation}
represents the coefficient at the fourth order nonlinear term caused by the quantum fluctuations
\cite{Ferrier-Barbut PRL 16}, \cite{Baillie 16}, \cite{Pfau Nature 16}, \cite{Adhikari LPL 17},
where the quantum fluctuations are related to the SRI and the dipole-dipole interaction,
and $\mu$ is the electric/magnetic dipole moment,
$\theta$ is an angle in the spherical coordinates $r$, $\theta$, $\varphi$,
$g=4\pi\hbar^{2} a_{s}/m$,
$\epsilon_{dd}=a_{dd}/a_{s}$,
$a_{dd}=m\mu_{0}\mu^{2}/12\pi\hbar^{2}$ is the dipole length \cite{Lahaye RPP 09}, \cite{Pfau Nature 16},
$a_{s}$ is the scattering length of the short range interaction,
$m$ is the mass of particle,
$\hbar$ is the Planck constant,
$\partial_{t}$ is the time derivative.
The dipolar part of the quantum fluctuations is found in Refs. \cite{Lima PRA 11}, \cite{Lima PRA 12}.


For the complete analysis of the Rosensweig instability in the rare-earth atoms,
the many-particle quantum hydrodynamics method is applied to the derivation of the model with anisotropic SRI.

This paper is organized as follows.
In Sec. II the formulation of basic ideas of the many-particle quantum hydrodynamics method is presented.
In Sec III the quantum stress tensor is calculated in the first order by the interaction radius.
In Sec. IV the quantum stress tensor is calculated in the third order by the interaction radius.
In Sec. V generalization of the Bogoliubov spectrum is demonstrated.
In Sec. VI the hydrodynamic equations are rewritten for the plane-like nonlinear objects propagating at the arbitrary angle to the preferable direction crated by the anisotropic interaction.
In Sec. VII is on a possibility of the reduction of the nonlocal nonlinearity to a local one.
In Sec. VIII models the bright solitons in the repulsive BEC appearing due to the nonlocal nonlinearities.
In Sec. IX a brief summary of obtained results is presented.

\section{On the derivation of hydrodynamic equations}

Derivation of the model is based on the many-particle Schrodinger equation containing the following Hamiltonian:
\begin{equation}\label{Hamiltonian micro}
\hat{H}=\sum_{i}\biggl(\frac{\hat{\textbf{p}}^{2}_{i}}{2m_{i}}+V_{ext}(\textbf{r}_{i},t)\biggr)
+\frac{1}{2}\sum_{i,j\neq i}U(\textbf{r}_{i}-\textbf{r}_{j}) ,\end{equation}
where $m_{i}$ is the mass of i-th particle, $\hat{\textbf{p}}_{i}=-\imath\hbar\nabla_{i}$ is the momentum of i-th particle,
$U_{ij}=U(\textbf{r}_{i}-\textbf{r}_{j})$ is the potential of interparticle interaction, $V_{ext}(\textbf{r}_{i},t)$ is the potential of external field acting on particles.

Derivation of the model starts with the definition of the simplest collective variable
which is the quantum concentration of particles.
It appears as the quantum mechanical average of corresponding operator:
\begin{equation}\label{concentr def}n(\textbf{r},t)=\int
dR\sum_{i}\delta(\textbf{r}-\textbf{r}_{i})\psi^{*}(R,t)\psi(R,t),\end{equation}
where $dR=\prod_{i=1}^{N}d\textbf{r}_{i}$ is the element of volume in $3N$ dimensional configurational space,
with $N$ is the number of particles.
Its evolution calculated with the application of the many-particle Schrodinger equation and Hamiltonian (\ref{Hamiltonian micro}) leads to the continuity equation: $\partial_{t}n+\nabla \textbf{j}=0$. The last one gives the explicit form of the current:
\begin{equation}\label{j def}
\textbf{j}(\textbf{r},t)=\int dR\sum_{i}\delta(\textbf{r}-\textbf{r}_{i}) \frac{1}{2m_{i}}(\psi^{*}(R,t)\hat{\textbf{p}}_{i}\psi(R,t)+c.c.).\end{equation}

Next, the evolution of current leads to the Euler equation \cite{Andreev PRA08}:
\begin{equation}\label{Euler v1} mn(\partial_{t}+\textbf{v}\cdot\nabla)v^{\alpha}+\partial_{\beta}(p^{\alpha\beta}+\sigma^{\alpha\beta}+T^{\alpha\beta}) =-n\partial^{\alpha}V_{ext} ,\end{equation}
where the current is represented via the velocity field $\textbf{j}(\textbf{r},t)=n(\textbf{r},t)\textbf{v}(\textbf{r},t)$,
$p^{\alpha\beta}$ is the thermal pressure caused by the local distribution of particles on states with non-zero energy:
the thermal effects or the quantum fluctuations,
$\sigma^{\alpha\beta}$ is the quantum stress tensor caused by the short range interaction (SRI),
and
$T^{\alpha\beta}= -(\hbar^{2}/4m)[\partial^{\alpha}\partial^{\beta}n-(\partial^{\alpha}n)(\partial^{\beta}n)/n]$
is the quantum Bohm potential
(this is a simplified form suitable for the BEC,
while the general form is presented by equation (17) in Ref. \cite{Andreev PRA08} and equation (10) in Ref. \cite{Andreev PRB 11}).
Tensor $\sigma^{\alpha\beta}$ appears at the calculation of the force field.
It is assumed that the masses of all particles are equal to each other
and written without subindex: $m$.
Originally, in the Euler equation, the force field appears as follows
\begin{equation} \label{ForceField} \textbf{F}(\textbf{r},t)=-\int dR\sum_{i,j\neq i}\delta(\textbf{r}-\textbf{r}_{i})(\nabla_{i}U(\textbf{r}_{ij}))\psi^{*}(R,t)\psi(R,t).\end{equation}
It can be symmetrized relatively pair of interacting particles
$$\textbf{F}(\textbf{r},t)=-\frac{1}{2}\int dR\sum_{i,j\neq i}[\delta(\textbf{r}-\textbf{r}_{i})-\delta(\textbf{r}-\textbf{r}_{j})]\times$$
\begin{equation} \label{ForceField symm} \times\nabla_{i}U(\textbf{r}_{ij})\cdot\psi^{*}(R,t)\psi(R,t) .\end{equation}
Next, introducing the coordinates of relative motion and center of mass for each pair of particles $\textbf{R}_{ij}=\frac{1}{2}(\textbf{r}_{i}+\textbf{r}_{j})$, $\textbf{r}_{ij}=\textbf{r}_{i}-\textbf{r}_{j}$, represent coordinates of $i$-th and $j$-th particles via $\textbf{r}_{ij}$ and $\textbf{R}_{ij}$ in equation (\ref{ForceField symm}).

The SRI is considered.
Hence, the potential $U_{ij}$ and its derivatives $\partial_{\alpha}U_{ij}$ go to zero for the large interparticle distances $\textbf{r}_{ij}$.
Therefore, equation (\ref{ForceField}) is nontrivial for the small values of $\textbf{r}_{ij}$.
Consequently, we can expand equation (\ref{ForceField}) in the Taylor series on parameter $\textbf{r}_{ij}$.
The $\delta$ functions and the $i$-th and $j$-th arguments of the wave functions are involved in the expansion.
The zeroth term and the even terms are equal to zero due to the symmetry of the equation (\ref{ForceField}) relatively $i$-th and $j$-th particles.
Hence, we deal with odd 
terms.
In the simplest case, we restrict our analysis with the first and third terms on the interaction radius.

This paper is a development of the many-particle quantum hydrodynamic method
which gives a representation of the many-particle quantum system described by the time-dependent Schrodinger equation in terms of the collective variables.
This method suggested in \cite{Maksimov QHM 99-01} for the quantum plasmas
is applied to derivation of the Gross-Pitaevskii equation and its nonlocal generalization in \cite{Andreev PRA08}.
It is done with accordance with the earlier proved possibility of such kind of derivation \cite{Erdos PRL 07}.
In this context, it is interesting to mention work \cite{Kronke PRA 18},
where the quantum dynamics of finite ultracold bosonic ensembles based on the Born-Bogoliubov-Green-Kirkwood-Yvon hierarchy
for equations of motion for few-particle reduced density matrices with the necessary truncation.
The multiconfiguration time-dependent Hartree method for bosons equations of motion of \cite{Alon PRA 08}
for the specific wave function ansatz as their starting point.

\section{Quantum stress tensor in the first order by the interaction radius}

Considering the quantum stress tensor (QST) in the first order by the interaction radius
find the following representation in terms of the many-particle wave function:
$$\sigma_{1}^{\alpha\beta}(\textbf{r},t)=-\frac{1}{2}\int dR'\sum_{i,j\neq i}\delta(\textbf{r}-\textbf{R}_{ij})r^{\beta}_{ij}\times$$
\begin{equation} \label{sigma in 1 or} \times\frac{\partial U(\textbf{r}_{ij})}{\partial r_{ij}^{\alpha}}\psi^{*}(R',t)\psi(R',t),\end{equation}
where $R'=\{\textbf{r}_{1}, ..., \textbf{R}_{ij}, ..., \textbf{R}_{ij}, ..., \textbf{r}_{N}\}$
with $\textbf{R}_{ij}$ is located on the $i$-th and $j$-th places, but $dR'=d\textbf{R}_{ij}d\textbf{r}_{ij}dR_{N-2}$, where $dR_{N-2}$ does not contain the contribution of the $i$-th and $j$-th particles.

The particles under consideration have a preferable direction resulting in the nonisotropic interaction.
Being in an arbitrary state the system consist of particles with the arbitrary orientation.
The summation in the quantum stress tensor on all pair of particles leads to the cancelation or at leats the decrease of the anisotropic part.
However, if all particles oriented in the same direction the anisotropic part is not reduced.
Moreover, the part of integral (\ref{sigma in 1 or}) and the QST in the higher orders containing $\textbf{r}_{ij}$ can be extracted as the common multiplier.
This assumption allows us to go to the simplified representation of the QST presented by the next equation.
Let us mention here that the considered particles posses the magnetic moment.
Therefore, orientation of the particles relatively the SRI is related to the creation of the spin polarization in the system.
In the simplest case, the preferable direction of the SRI coincides with the spin direction.
If it has an angle with spin  the spin polarization create a partial orientation.
Hence the perpendicular parts would cancel each other,
but the part of the anisotropic SRI parallel to the spin is not zero and contribute in the collective interaction.
This part enters the model developed in this paper.

The quantum stress tensor can be represented in a form containing two-particle concentration
\begin{equation} \label{sigma in 1 or hz n2} \sigma_{1}^{\alpha\beta}(\textbf{r},t)= -\frac{1}{2}Tr(n_{2}(\textbf{r},\textbf{r}',t)) \int
r^{\beta}\frac{\partial U(\textbf{r})}{\partial r^{\alpha}}d\textbf{r} ,\end{equation}
where
$Tr n_{2}(\textbf{r},\textbf{r}')=n_{2}(\textbf{r},\textbf{r})$
is the trace of the function of two arguments, and
\begin{equation} \label{n2 def}n_{2}(\textbf{r},\textbf{r}',t)=\int dR\sum_{i,j\neq i}\delta(\textbf{r}-\textbf{r}_{i})\delta(\textbf{r}'-\textbf{r}_{j})\psi^{*}(R,t)\psi(R,t) \end{equation}
is the explicit form of two-particle concentration.

Considering even and axial symmetric anisotropic potentials, include the expansion of the potential on the spherical functions
\begin{equation}\label{expansion of U on Y} U(r,\theta)=\sqrt{4\pi}\sum_{l=0}^{\infty}Y_{2l,0}(\theta)U_{2l}(r)\end{equation}
find a generalization of the quantum stress tensor obtained in \cite{Andreev PRA08}
$$\sigma_{1,BEC}^{\alpha\beta}=-\sqrt{\pi}Tr(n_{2}(\textbf{r},\textbf{r}',t)) \sum_{l=0}^{\infty}$$
\begin{equation}\label{sigma n_2+gen coef} \int d\textbf{r}  \biggl(Y_{2l,0}(\theta)\frac{r^{\alpha}r^{\beta}}{r}\frac{\partial U_{2l}(r)}{\partial r}
+ U_{2l}(r)\frac{e_{\theta}^{\alpha}r^{\beta}}{r}\frac{\partial Y_{2l,0}(\theta)}{\partial \theta}\biggr),\end{equation}
where $\textbf{e}_{\theta}$ is the unit vector in the spherical coordinate system $r$, $\theta$, $\varphi$.
Presence of $r^{\alpha}r^{\beta}$ and $e_{\theta}^{\alpha}r^{\beta}$ under the integral shows
that two terms from expansion (\ref{expansion of U on Y}) give a contribution.
These are terms proportional $Y_{00}(\theta)$ and $Y_{20}(\theta)$. The second of them brings anisotropy in the model.
However, the calculation of the tensor structure of the quantum stress tensor
in the first order by the interaction radius shows cancelation of the term proportional to $Y_{20}(\theta)$.
Hence, the anisotropy gives no contribution to the quantum stress tensor
in the first order by the interaction radius even if anisotropic part of the potential is comparable with the isotropic one.

The two-particle concentration is calculated for the Bose-Einstein condensate in \cite{Andreev PRA08},
where it is found that $Tr(n_{2}(\textbf{r},\textbf{r}',t))=n^{2}(\textbf{r},t)$.

After all, obtain the standard result for the quantum stress tensor in the first order by the interaction radius:
\begin{equation}\label{sigma BEC via wf} \sigma_{1,BEC}^{\alpha\beta}=\frac{1}{2}g\delta^{\alpha\beta} n^{2},\end{equation}
where $g=\int U_{0}(r) d\textbf{r}$.

Substituting the found QST in the Euler equation,
find the following result
$$ mn(\partial_{t} +\textbf{v}\cdot\nabla)v^{\alpha}
-\frac{\hbar^{2}}{2m}n\partial^{\alpha}\frac{\triangle\sqrt{n}}{\sqrt{n}}$$
\begin{equation}\label{Euler v2}  +g n\partial^{\alpha}n =-n\partial^{\alpha}V_{ext},\end{equation}
where $p^{\alpha\beta}=0$ since bosons at the zero temperature are considered.

The Euler equation (\ref{Euler v2}) appears together with the continuity equation presented above.
Here, we represent the continuity equation via the velocity field:
\begin{equation}\label{cont eq via vel} \partial_{t}n+\nabla\cdot (n\textbf{v})=0. \end{equation}

Equations (\ref{Euler v2}) and (\ref{cont eq via vel}) correspond to the traditional Gross-Pitaevskii equation \cite{Dalfovo RMP 99}:
\begin{equation}\label{QBT GP first appearence} \imath\hbar\partial_{t}\Psi
=\biggl(-\frac{\hbar^{2}\nabla^{2}}{2m}+V_{ext} +g\mid\Psi\mid^{2}\biggr)\Psi,\end{equation}
with
\begin{equation}\label{def Phi} \Psi(\textbf{r},t)=\sqrt{n} e^{\imath m\phi/\hbar},\end{equation}
where $\textbf{v}=\nabla\phi$.

\section{Quantum stress tensor in the third order on the interaction radius}

The quantum hydrodynamic model of BEC of neutral atoms with the SRI considered up to the TOIR is developed in 2008 \cite{Andreev PRA08} for the isotropic potentials.
Here it is considered for the anisotropic potentials.

An analog of equation (\ref{sigma in 1 or})
obtained in the TOIR is a large equation. 
Majority of the terms in this equation do not contribute in the BEC dynamics being related to the excited states (see Appendix A).
For the BEC it simplifies to
$$\sigma^{\alpha\beta}_{3,BEC}(\textbf{r},t)$$
\begin{equation} \label{sigma in 3 or hz n2} =
-\frac{1}{48}\partial_{\gamma}\partial_{\delta}Tr(n_{2}(\textbf{r},\textbf{r}',t)) \cdot \int
r^{\beta}r^{\gamma}r^{\delta}\frac{\partial U(\textbf{r})}{\partial r^{\alpha}}d\textbf{r}.\end{equation}

Calculation of the tensor structure of the QST in the TOIR gives the following result
\begin{equation} \label{tensor structure in 3 order} \int
r^{\beta}r^{\gamma}r^{\delta}\frac{\partial U(\textbf{r})}{\partial r^{\alpha}}d\textbf{r}=
-\frac{1}{3}\tilde{g}_{2,0}I_{0}^{\alpha\beta\gamma\delta}
+\frac{1}{\sqrt{5}}\tilde{g}_{2,2}I_{2}^{\alpha\beta\gamma\delta},\end{equation}
where
$I_{0}^{\alpha\beta\gamma\delta}=\delta^{\alpha\beta}\delta^{\gamma\delta}+\delta^{\alpha\gamma}\delta^{\beta\delta}+\delta^{\alpha\delta}\delta^{\beta\gamma}$ is a symmetric tensor existing in the term describing isotropic part,
and
$I_{2}^{\alpha\beta\gamma\delta}$ is a nonsymmetric tensor, or more precisely, it is symmetric relatively permutations of the last three indexes, but it is not symmetric relatively permutations of the first index and other indexes. Tensor $I_{2}^{\alpha\beta\gamma\delta}$ has the following elements:  $I_{2}^{xxxx}=I_{2}^{yyyy}=1$, $I_{2}^{zzzz}=-2$, $I_{2}^{xxzz}=I_{2}^{yyzz}=-2/3$, $I_{2}^{xxyy}=I_{2}^{yyxx}=I_{2}^{zzxx}=I_{2}^{zzyy}=1/3$ and allowed permutations of indexes, other elements are equal to zero. Equation (\ref{sigma in 3 or hz n2}) contains two interaction constants $g_{2,l}=\int r^{2}U_{l}(r)d\textbf{r}/24$, $g_{2,l}=\tilde{g}_{2,l}/24$, where $l=0$, $2$.

Present the Euler equation with the SRI considered up to the TOIR
$$ mn(\partial_{t} +\textbf{v}\cdot\nabla)v^{\alpha} +n\partial^{\alpha}V_{ext} -\frac{\hbar^{2}}{2m}n\partial^{\alpha}\frac{\triangle\sqrt{n}}{\sqrt{n}}$$
\begin{equation}\label{Euler v3} +g n\partial^{\alpha}n+\frac{g_{2,0}}{2}\partial^{\alpha}\triangle n^{2}
-\frac{g_{2,2}}{2\sqrt{5}}  I_{2}^{\alpha\beta\gamma\delta} \partial_{\beta}\partial_{\gamma}\partial_{\delta}n^{2}=0.\end{equation}

Obviously, the found generalized Euler equation corresponds to an approximation generalizing the Gross-Pitaevskii equation.

Interaction constants $g$ and $g_{2,0}$ are defined by $U_{0}$.
Hence, they contributions $g\partial_{\alpha}n^{2}/2$ and $g_{2,0}\partial_{\alpha}\triangle n^{2}/2$ can be compared.
Let us introduce the radius of interaction $r_{0}$
in a way that the interparticle interaction $U_{ij}(\textbf{r})$ or at least $U_{0}(r)$ is negligible for $r > r_{0}$.
Consider dimensionally reduced parameters $\hat{g}$ and $\hat{g}_{2,0}$
introduced as follows $g=r_{0}^{3}\hat{g}$ and $g_{2,0}=r_{0}^{5}\hat{g}_{2,0}$,
so $[\hat{g}]=[\hat{g}_{2,0}]=erg$.
We have $\hat{g}=\int U_{0}(r_{0}\xi) d^{3}\xi$ and $\hat{g}_{2,0}=\int\xi^{2}U_{0}(r_{0}\xi) d^{3}\xi$,
where $\xi=r/r_{0}$ and $\xi<1$ for the area of the nontrivial values of the potential $U_{0}$.
Therefore, the presence of $\xi^{2}$ under the integral in the definition of $\hat{g}_{2,0}$ decreases its value in compare with $\hat{g}$.
It shows $\hat{g}_{2,0}\ll\hat{g}$.
Our comparison shows that the first order by the interaction radius dominates over higher orders.
However, interaction constant $g_{2,2}$ is defined by another part of the anisotropic potential $U_{2}$.
Therefore, constants $g_{2,0}$ and $g_{2,2}$ can have different relationships $g_{2,0}>g_{2,2}$ or $g_{2,0}<g_{2,2}$.
If function $U_{2}$ is large in compare with $U_{0}$ (the large dipole anisotropy limit) the parameter $\hat{g}_{2,2}=\int\xi^{2}U_{2}(r_{0}\xi) d^{3}\xi$ can be large in compare with $\hat{g}_{2,0}$ and comparable with $\hat{g}$.


\begin{figure}
\includegraphics[width=8cm,angle=0]{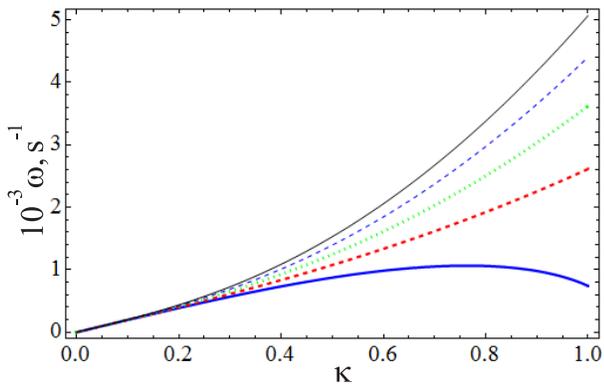}
\caption{\label{SUSD ObEx 01}
The Bogiliubov spectrum for different direction of wave propagation relatively preferable direction created by the anisotropy of SRI.
Increase of angle from $0$ to $\pi/2$ increases the frequency.
The frequency difference becomes noticeable in the short-wavelength limit.
The scattering length is chosen to be $a=10^{-7}$ cm
with $n_{0}=10^{14}$ cm$^{-3}$, the mass $m=167$ aem, 
$k=\kappa\sqrt[3]{n_{0}}$, $G_{2,0}=0.2$, $G_{2,2}=0.5$. }
\end{figure}

\begin{figure}
\includegraphics[width=8cm,angle=0]{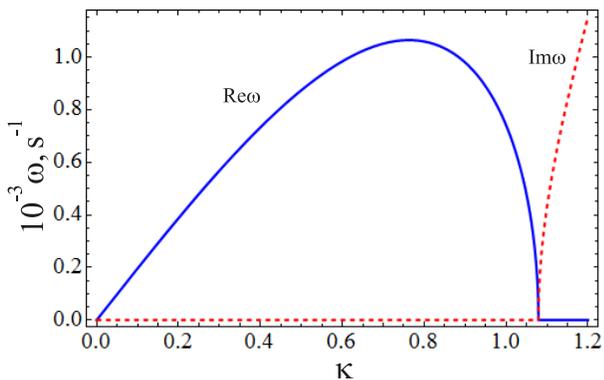}
\caption{\label{SUSD ObEx 02} The small angle short-wavelength instability is demonstrated for the large enough interaction constants of the TOIR.}
\end{figure}

\section{Collective excitations}

Spectrum of the collective excitations propagating as the plane wave in the infinite uniform BEC generalizing the Bogoliubov spectrum is
\begin{equation}\label{spectrum} \omega^{2}= \frac{n_{0}}{m} \biggl(gk^{2}+\frac{\hbar^{2}k^{4}}{4mn_{0}} -g_{2,0}k^{4}-\frac{g_{2,2}}{\sqrt{5}}k^{4}(3\cos^{2}\theta-1)\biggr). \end{equation}
The first term on the right-hand side is the short-range interaction
calculated in the first order by the interaction radius existing in the original Bogoliubov spectrum.
The second term is the quantum Bohm potential contribution.
The third and fourth terms are caused by the TOIR.
The third term is related to the isotropic part of the TOIR.
The last term in this equation is caused by the anisotropic part of the short range interaction.
The third and fourth terms give a generalization of the Bogoliubov spectrum caused by the short range interaction in the third order by the interaction radius.

Equation (\ref{spectrum}) is found from the hydrodynamic equations (\ref{cont eq via vel}) and (\ref{Euler v3}).
Derivation of equation (\ref{spectrum}) is made in the linear approximation
on the small perturbations of the hydrodynamic functions relatively to the equilibrium state $n_{0}$, and $\textbf{v}_{0}=0$.
Hence, the hydrodynamic variables are expressed as follows: $n=n_{0}+\delta n$, and $\textbf{v}=0+\delta\textbf{v}$.
The perturbations are presented in the following form:
$\delta n=N\exp(-\imath\omega t+\imath k_{x}x+\imath k_{z}z)$,
and $\delta\textbf{v}=V\exp(-\imath\omega t+\imath k_{x}x+\imath k_{z}z)$,
where $N$ and $V$ are the amplitudes of the perturbations.

The terms proportional to $k^{4}$ dominate in the short-wavelength limit
(the last three terms in equation (\ref{spectrum})).
The second term is positive.
The third and fourth terms can be negative and they can overcome the second term.
Hence, the short-wavelength instability can occur in the BEC due to the interaction in the third order by the interaction radius.

Presented comparison of three terms demonstrates
that the parameter $\hbar^{2}/4mn_{0}$ is the natural unit for the interaction constants $g_{2,0}$ and $g_{2,2}$.
Introduce corresponding dimensionless interaction constants $G_{2,0}=4mn_{0}g_{2,0}/\hbar^{2}$ and $G_{2,2}=4mn_{0}g_{2,2}/\hbar^{2}$.

Focus on the isotropic regime $g_{2,2}=0$.
The repulsive SRI $g>0$ leads to the long-wavelength stability of the Bogoliubov spectrum.
However, short-wavelength instability can exist
if $G_{2,0}$ is larger then the quantum Bohm potential contribution $1$,
where $G_{2,0}>0$ for the repulsive interaction
since interaction constants $g$ and $G_{2,0}$ are the moments of the same partial potential $U_{0}$.
If $1>G_{2,0}>0$ the spectrum (\ref{spectrum})) is stable,
but the frequency $\omega$ decreases in the area of large $k$ in compare with the first order by the interaction radius approximation.

Interaction constant $g_{2,2}$ is the moment of the second partial potential $U_{2}$.
Therefore, it does not have direct relation to $g_{2,0}$.
Hence, it is possible to have positive or negative $g_{2,2}$ at the positive $g_{2,0}$.
Anyway, angle dependence of the last term in equation (\ref{spectrum}) gives different signs of the last term depending on the wave propagation direction.

Negative contribution of the TOIR in the square of frequency is demonstrated in Fig. \ref{SUSD ObEx 01}.
Changing direction of wave propagation we change  the relative contribution of the TOIR.
If angle is small enough and the TOIR interaction constants are large enough the frequency get the zero value at some wave vector $\kappa>1$.
Spectrum becomes imaginary and condition for an instability occurs (see Fig. \ref{SUSD ObEx 02}).

The instability can be caused by the isotropic part of the TOIR. 
However, in accordance with the described estimations, 
the anisotropic interaction constant can be larger then the isotropic TOIR constant.
Therefore, the instability is easier detectable in the BEC of atoms with the anisotropic interaction.

\section{Propagation of the nonlinear perturbations in an arbitrary direction}

Consider the wave propagation in the direction $h$ which has angle $\theta$ with the anisotropy axis (the $z$-axis).
Change of coordinate in $h$ direction corresponds to the change of $x$ and $z$
(it is assumed that $y=0$) in accordance with the following relation $h=x\sin\theta+z\cos\theta$.
The velocity field perturbations in the direction $h$ has corresponding form $v_{h}=v_{x}\sin\theta+v_{z}\cos\theta$.


The continuity equation transforms in the following way for the perturbations propagating in direction $h$
\begin{equation}\label{} \partial_{t}n +\nabla\cdot(n\textbf{v})=0, \end{equation}
\begin{equation}\label{} \partial_{t}n +\partial_{x}(nv_x) +\partial_{z}(nv_z)=0, \end{equation}
\begin{equation}\label{} \partial_{t}n +\sin\theta(nv_x)' +\cos\theta(nv_z)'=0, \end{equation}
and
\begin{equation}\label{continuity h} \partial_{t}n +(n v_h)'=0, \end{equation}
since $\partial_{x}f=f' \sin\theta$, and $\partial_{z}f=f' \cos\theta$, where $f=f(h)$ an arbitrary hydrodynamic function $n$, $v_x$, $v_z$ or their combination as a function of coordinate $h$, and $f'=df/dh$.

Corresponding modification is found for the $x-$ and $z-$projections of the Euler equation
$$mn(\partial_{t}v_{x}+v_{h} v_{x}')+n\sin\theta V_{ext}'-\frac{\hbar^{2}}{2m}\sin\theta n \biggl[\frac{(\sqrt{n})''}{\sqrt{n}}\biggr]'$$
$$+gn\sin\theta n'+ \frac{1}{2}g_{2,0}\sin\theta (n^2)'''$$
\begin{equation}\label{Euler x via h}
-\frac{g_{2,2}}{2\sqrt{5}} \sin\theta \biggl(\sin^{2}\theta-\frac{2}{3}\cos^{2}\theta\biggr) (n^2)'''=0, \end{equation}
and
$$mn(\partial_{t}v_{z}+v_{h} v_{z}')+n\cos\theta V_{ext}'-\frac{\hbar^{2}}{2m}\cos\theta n \biggl[\frac{(\sqrt{n})''}{\sqrt{n}}\biggr]'$$
$$+gn\cos\theta n'+ \frac{1}{2}g_{2,0}\cos\theta (n^2)'''$$
\begin{equation}\label{Euler z via h}
-\frac{g_{2,2}}{2\sqrt{5}} \cos\theta \biggl(\frac{1}{3}\sin^{2}\theta-2\cos^{2}\theta\biggr) (n^2)'''=0. \end{equation}
The $y-$ projection of the Euler equation is not involved.

Multiply equation (\ref{Euler x via h}) on $\sin\theta$ and equation (\ref{Euler z via h}) on $\cos\theta$ and combine obtained equations to get equation for $\partial_{t}v_{h}$:

$$mn(\partial_{t}v_{h}+v_{h} v_{h}')+n V_{ext}'$$
$$-\frac{\hbar^{2}}{2m} n \biggl[\frac{(\sqrt{n})''}{\sqrt{n}}\biggr]' +gn n' +\frac{g_{2,0}}{2}(n^2)''' $$
\begin{equation}\label{Euler h}
-\frac{g_{2,2}}{2\sqrt{5}} \biggl[\sin^{4}\theta -\frac{1}{3}\sin^{2}\theta\cos^{2}\theta -2\cos^{4}\theta\biggr](n^2)''' =0. \end{equation}

Equation (\ref{Euler h}) shows that the third order on the interaction radius allows to introduce a single effective constant combining isotropic and nonisotropic parts
\begin{equation}\label{g 2 def}
g_{2}\equiv g_{2,0}-\frac{g_{2,2}}{\sqrt{5}} \biggl[\sin^{4}\theta -\frac{1}{3}\sin^{2}\theta\cos^{2}\theta -2\cos^{4}\theta\biggr]. \end{equation}
Hence, the last two terms in equation (\ref{Euler h}) can be written in the following way $\frac{g_{2}}{2}(n^2)'''$.

\begin{figure}
\includegraphics[width=8cm,angle=0]{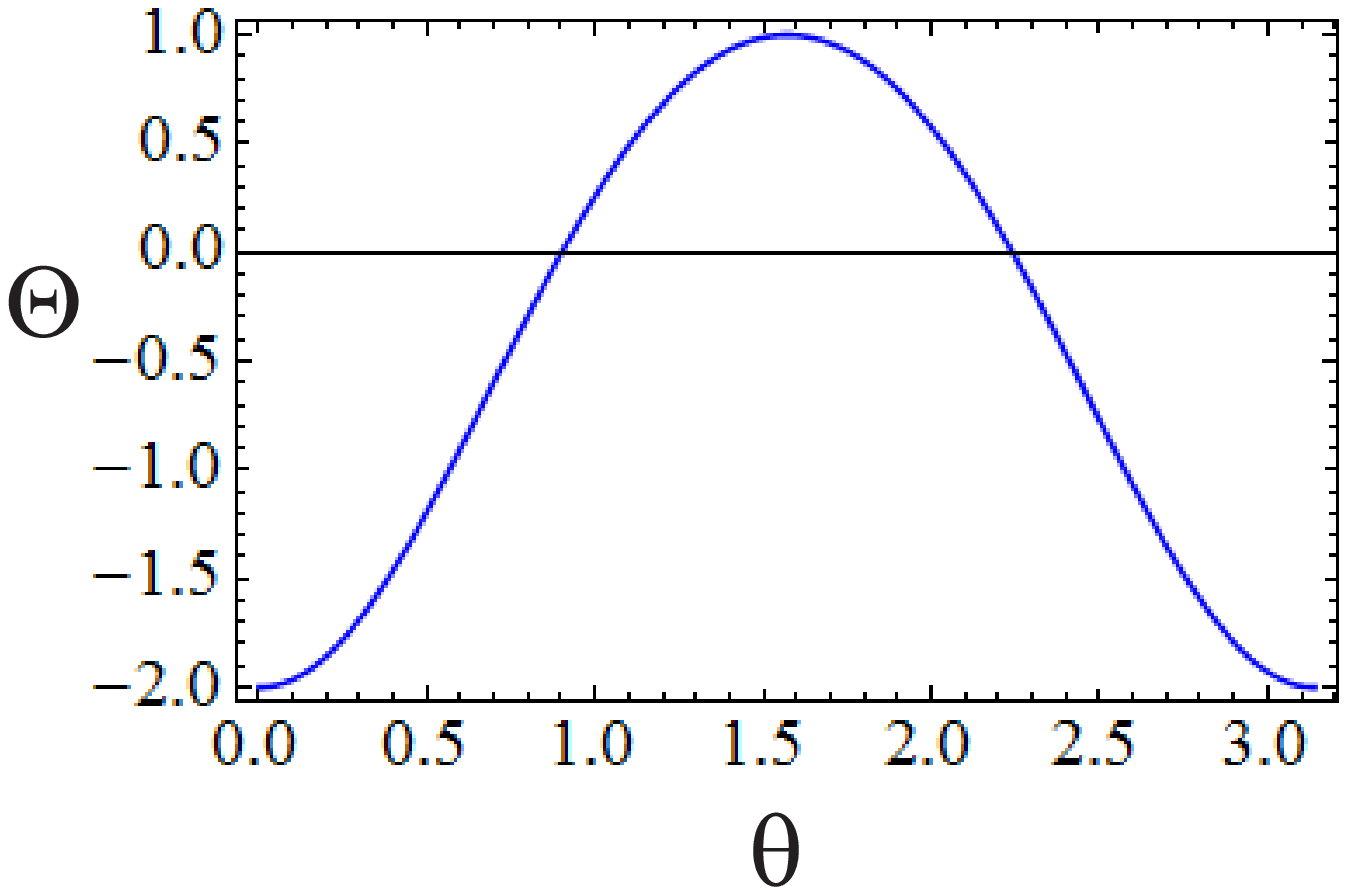}
\includegraphics[width=8cm,angle=0]{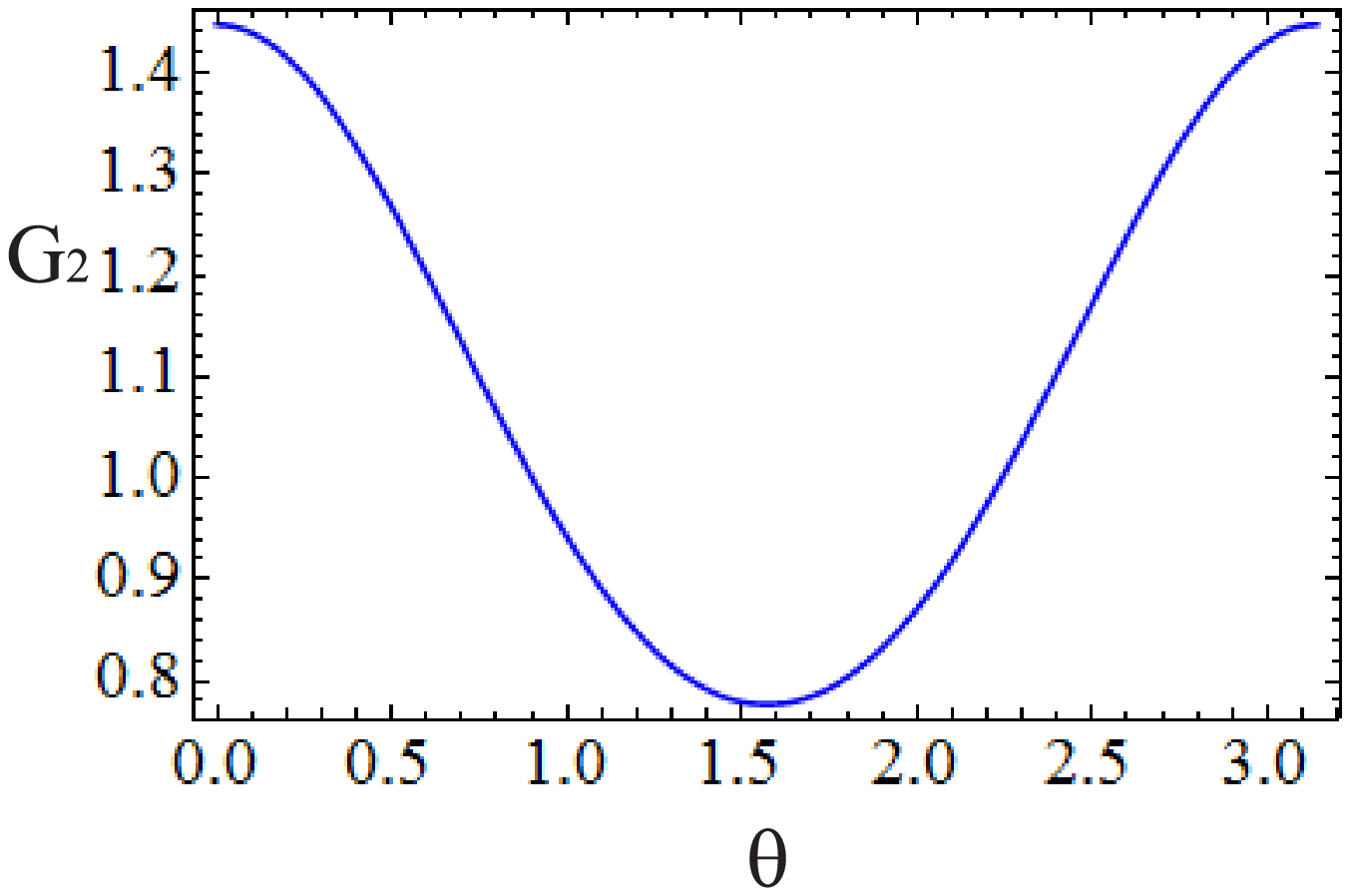}
\caption{\label{SUSD ObEx 03}  
The figure shows
1. $\Theta\equiv\sin^{4}\theta -\frac{1}{3}\sin^{2}\theta\cos^{2}\theta -2\cos^{4}\theta$ which is located in front of $g_{2,2}$;
2. Parameter $G_{2}\equiv4mn_{0}g_2/\hbar^{2}$ at $G_{2,0}=1$ and $G_{2,2}=0.1$.}
\end{figure}

\begin{figure}
\includegraphics[width=8cm,angle=0]{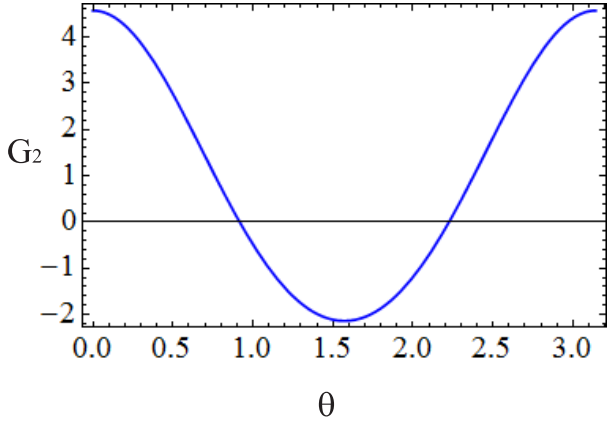}
\caption{\label{SUSD ObEx 04}
The figure shows a regime opposite to Fig. (\ref{SUSD ObEx 01}), where $g_{2,0}<g_{2,2}$
It demonstrates parameter $G_2$ at $G_{2,0}=0.1$ and $G_{2,2}=1$.}
\end{figure}

\begin{figure}
\includegraphics[width=8cm,angle=0]{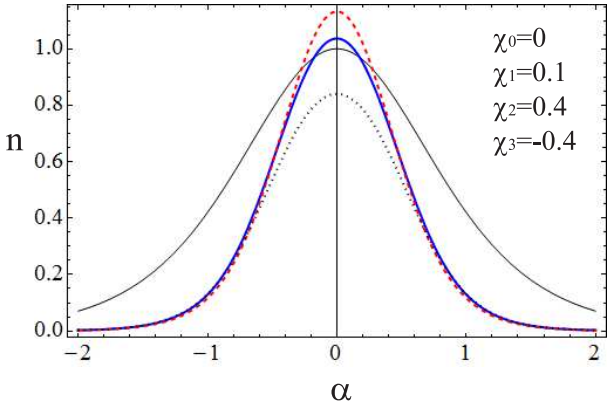}
\caption{\label{SUSD ObEx 05} 
Modification of the bright soliton form under the influence of the interaction in the TOIR approximation. 
Thin continuous line shows bright soliton solution in the Gross-Pitaevskii approximation. 
The thick continuous line is made for $\chi_{1}=0.1$. 
The dotted line is calculated for the larger values of $\chi_{2}=0.4$. 
The dashed (upper) curve is obtained for the negative value of $\chi$.}
\end{figure}

Dimensionless form of parameter $g_{2}$ is presented in Fig. \ref{SUSD ObEx 03} and \ref{SUSD ObEx 04} for chosen values of $g_{2,0}$ and $g_{2,2}$. 
It is found that parameter $g_{2}$ can change the sign. 
Hence, the repulsion existing at small angles and positive $g_{2,i}$ 
can be changed by the attraction for the large enough $g_{2,2}$ (see Fig. \ref{SUSD ObEx 04}).

\section{On approximate reduction of the nonlocal nonlinearity to the quintic nonlinearity and the bright soliton in the attractive BEC}

Here we apply a method of nonperturbative analysis of the excitations in BEC.
It allows to consider the large amplitude bright soliton in the one dimensional attractive BEC.
This method is a nonperturbative on the amplitude of the excitations.
However, below, we explicitly use the condition of the weak interaction in the TOIR approximation in compare with the interaction in the Gross-Pitaevskii approximation (or the FOIR approximation).

\subsection{General analysis of 1D regime}

It has been shown that the nonlocal nonlinearity existing in the third order by the interaction radius in the isotropic regime proportional to $\partial^{\alpha}\triangle n^{2}$ can be approximately reduced to the local form.
It can be represented as a nonlinearity of fifth degree \cite{Andreev Izv.Vuzov. 10}.
Hence, the Euler equation contain the following force field $\beta\partial^{\alpha} n^{3}$.
This reduction is not a general property, but it approximately happens for the bright soliton structure.

Consider the one dimensional regime of the quantum hydrodynamic equations, for the arbitrary direction
\begin{equation}\label{Euler v4 1D} mn(\partial_{t} +v_{h}\partial_{h})v_{h} =\frac{\hbar^{2}}{2m}n\partial_{h}\frac{\partial_{h}^{2}\sqrt{n}}{\sqrt{n}}
-g n\partial_{h}n-\frac{g_{2}}{2}\partial_{h}^{3} n^{2}. \end{equation}

Divide equation (\ref{Euler v4 1D}) by $mn$ and integrate it over $h$ to find
\begin{equation}\label{Euler v4 1D integrated}  \partial_{t}\int v_{h}dh +\frac{1}{2}v_{h}^{2} =\frac{\hbar^{2}}{2m^{2}}\frac{\partial_{h}^{2}\sqrt{n}}{\sqrt{n}} -\frac{g}{m} n -\frac{g_{2}}{2m}\int\frac{\partial_{h}^{3} n^{2}}{n}dh. \end{equation}

Next, multiply equation (\ref{Euler v4 1D integrated}) by $\sqrt{n} \partial_{h}\sqrt{n}$ and obtain the following
$$ \sqrt{n} \partial_{h}\sqrt{n}\cdot\partial_{t}\int v_{h}dh +\frac{1}{2}\sqrt{n} \partial_{h}\sqrt{n}\cdot v_{h}^{2}
=\frac{\hbar^{2}}{2m^{2}} \partial_{h}\sqrt{n}\cdot\partial_{h}^{2}\sqrt{n}$$
\begin{equation}\label{Euler v4 1D integrated and multiplied}
-\frac{g}{m} \sqrt{n}n \partial_{h}\sqrt{n} -\frac{g_{2}}{2m} \sqrt{n} \partial_{h}\sqrt{n}\cdot \int\frac{\partial_{h}^{3} n^{2}}{n}dh. \end{equation}

We continue with the representation of the one dimensional hydrodynamic equations following Ref. \cite{Fedele EPJB 02}.
Next, we modify the continuity equation.
First, the continuity equation is multiplied by the velocity and splitted in the following way
$nv_{h}\partial_{h}v_{h}=-v_{h}\partial_{t}n-v_{h}^{2}\partial_{h}n$.
Next, add the following term in both sides of the presented equations $n\partial_{t}v_{h}$.
The left-hand side of the obtained equation
$n(\partial_{t}v_{h}+v_{h}\partial_{h}v_{h})=-v_{h}\partial_{t}n-v_{h}^{2}\partial_{h}n+n\partial_{t}v_{h}$
coincides with the kinematic part of the Euler equation placed on the left-hand side of equation (\ref{Euler v4 1D}).
Hence, the corresponding substitution is made in the Euler equation:
$$ m(n\partial_{t}v_{h}-v_{h}\partial_{t}n-v_{h}^{2}\partial_{h}n) =\frac{\hbar^{2}}{2m}n\partial_{h}\frac{\partial_{h}^{2}\sqrt{n}}{\sqrt{n}}$$
\begin{equation}\label{Euler v5 1D} -g n\partial_{h}n-\frac{g_{2}}{2}\partial_{h}^{3} n^{2}. \end{equation}

The quantum Bohm potential needs to be rewritten for further decompositions:
$\partial_{h}\frac{\partial_{h}^{2}\sqrt{n}}{\sqrt{n}}$
$=\frac{1}{n}(\partial_{h}^{3} n/2 -4\partial_{h}\sqrt{n} \cdot\partial_{h}^{2}\sqrt{n})$.

Obtained equation
$$ m(n\partial_{t}v_{h}-v_{h}\partial_{t}n-v_{h}^{2}\partial_{h}n)
=\frac{\hbar^{2}}{4m}\partial_{h}^{3} n $$
\begin{equation}\label{Euler v5 1D} -2\frac{\hbar^{2}}{m}\partial_{h}\sqrt{n}\cdot \partial_{h}^{2}\sqrt{n}
-g n\partial_{h}n-\frac{g_{2}}{2}\partial_{h}^{3} n^{2}. \end{equation}
allows to represent $-2\hbar^{2}\partial_{h}\sqrt{n} \cdot\partial_{h}^{2}\sqrt{n}/m$
via other terms and substitute it in equation (\ref{Euler v4 1D integrated and multiplied}) to find a required intermediate result:
$$ -2\partial_{h}n \cdot\partial_{t}\int v_{h}dh -n\partial_{t}v_{h} +v_{h}\partial_{t}n +\frac{\hbar^{2}}{4m^{2}}\partial_{h}^{3} n $$
\begin{equation}\label{Euler v5 1D combined} -\frac{3}{2}\frac{g}{m} n\partial_{h}n
-\frac{g_{2}}{m}\biggl[\frac{1}{2}\partial_{h}^{3} n^{2} +\partial_{h}n \cdot\int\frac{\partial_{h}^{3} n^{2}}{n}dh\biggr]=0.\end{equation}

Next, we consider a simplified regime 
which allows to describe the bright soliton in BEC and the reduction of the nonlocal nonlinearity existing in this regime.

The velocity field in BEC is the potential field $v_{h}=\partial_{h}\Theta$.
The nonzero value of the velocity is related to the solitons propagation moving with a constant velocity $v_{h}=v_{0}$.
It leads to the following form of the velocity potential $\Theta=v_{0}h-c_{0}t$,
where $c_{0}$ is a constant
which is not depended on space $h$ and time $t$.

It is necessary to find a solution of hydrodynamic equations in the stationary wave form.
Hence, the independent hydrodynamic function (the concentration) is assumed to be functions of $\eta=h-v_{0}t$.

Overall, we have that derivatives of the velocity in equation (\ref{Euler v5 1D combined}) go to zero
and $\partial_{h}=\partial_{\eta}$, $\partial_{t}=-v_{0}\partial_{\eta}$.
Hence, equation (\ref{Euler v5 1D combined}) simplifies in this regime to the following equation
$$2E_{0}\partial_{\eta}n +\frac{\hbar^{2}}{4m^{2}}\partial_{\eta}^{3}n-\frac{3g}{m}n\partial_{\eta}n$$
\begin{equation}\label{KdV eq in the third order int form}
-\frac{g_{2}}{m}\biggl(\partial_{\eta}n\cdot\int \frac{\partial_{\eta}^{3}n^{2}}{n}d\eta
+\frac{1}{2}\partial_{\eta}^{3}n^{2}\biggr)=0, \end{equation}
where parameter $E_{0}=c_{0}-v_{0}^{2}/2$ is introduced.

Restrict the analysis by the first order on the interaction radius and find the following simplified equation
\begin{equation}\label{KdV eq in the first order} 2E_{0}\partial_{\eta}n +\frac{\hbar^{2}}{4m^{2}}\partial_{\eta}^{3}n-\frac{3g}{m}n\partial_{\eta}n=0 \end{equation}
which gives the traditional bright soliton solution for the attractive BEC $g<0$.

The bright soliton solution appear from equation (\ref{KdV eq in the first order})
for the negative energy $E_{0}<0$ and the following boundary conditions
$\lim_{\eta\rightarrow\pm\infty}n(\eta)=0$.

Solution has the traditional form
\begin{equation}\label{KdV eq in the first order sol}
n(\eta)=\frac{n_{0}}{\cosh^{2}\biggl(\frac{m\sqrt{2\mid E_{0}\mid} \eta}{\hbar}\biggr)}, \end{equation}
where $n_{0}=2m\mid E_{0}\mid/\mid g\mid$.

\subsection{Isotropic case}

In the isotropic case the interaction constant $g_{2}$ reduces to $g_{2,0}$.

Assume that the third order by the interaction radius contribution
is significantly smaller then the GP contribution.
Hence, solve equation (\ref{KdV eq in the third order int form}) by the iteration method.
Equation (\ref{KdV eq in the first order}) and solution (\ref{KdV eq in the first order sol}) appear in the first iteration step.
For the second step, substitute solution (\ref{KdV eq in the first order sol})
in the third order by the interaction radius term in equation (\ref{KdV eq in the third order int form})
and, after integration over $\eta$, find the following result
$$2E_{0}n +\frac{\hbar^{2}}{4m^{2}}\partial_{\eta}^{2}n-\frac{3g}{2m}n^{2}$$
\begin{equation}\label{KdV eq in the third order appr form}
-\frac{192g_{2}m^{4}\mid E_{0}\mid^{3}}{\hbar^{2}g^{2}}\times\biggl[\cosh^{-4}\alpha -\frac{5}{6}\cosh^{-6}\alpha\biggr]=0, \end{equation}
where $\alpha\equiv \sqrt{2\mid E_{0}\mid} m\eta/\hbar$.

Next, represent the last term in equation (\ref{KdV eq in the third order appr form}) via the concentration
in accordance with equation (\ref{KdV eq in the first order sol}).
As the result, we find
\begin{equation}\label{KdV eq in the third order second iter via n} 2E_{0}n +\frac{\hbar^{2}}{4m^{2}}\partial_{\eta}^{2}n -\biggl(\frac{3g}{2m}-\frac{48m^{2}E_{0}g_{2}}{\hbar^{2}}\biggr)n^{2} -\frac{20mgg_{2}}{\hbar^{2}}n^{3}=0. \end{equation}

Equation (\ref{KdV eq in the third order second iter via n}) corresponds to the following effective NLSE:
\begin{equation}\label{NLSE eq with quintic nonlin} \imath\hbar\partial_{t}\Psi=-\frac{\hbar^{2}}{2m}\triangle\Psi +\tilde{g}\mid\Psi\mid^{2}\Psi +g_{f}\mid\Psi\mid^{4}\Psi, \end{equation}
where $\tilde{g}=g-32m^{3}E_{0}g_{2}/\hbar^{2}$, and $g_{f}=20m^{2}gg_{2}/\hbar^{2}$ is the interaction constant for the effective fifth order nonlinearity.

Approximate analysis of the nonlocal nonlinearity for the bright solitons shows
the reduction of the nonlocal nonlinearity to the quintic nonlinearity presented by the last term in equation (\ref{NLSE eq with quintic nonlin}).
It also causes the modification of the cubic nonlinearity via the shift of the interaction constant presented by $\tilde{g}$.

Equation (\ref{KdV eq in the third order second iter via n}) has an analytical soliton solution
\begin{equation}\label{KdV eq in the third order sol} n(\eta)=\frac{2n_{0}}{\sqrt{\chi^{2}+0.5\chi+1}\cosh^{2}(2\alpha)-\chi+1}, \end{equation}
where 
$\chi=32m^{2}\mid E_{0}\mid g_{2}/\hbar^{2}g$.
The amplitude of the soliton is changed by the interaction in the TOIR approximation $\chi\neq0$.
For the isotropic case we can expect that interaction constants $g$ and $g_{2,0}$ have same sign.
Consequently, the ratio $g_{2}/g$ existing in $\chi$ is positive.
Hence, we have $\chi>0$.
It implies the increase of the soliton amplitude $n_{max}=n_{0}(1+3\chi/8)$.

Comparison of solutions (\ref{KdV eq in the first order sol}) and (\ref{KdV eq in the third order sol}) is presented in Fig. \ref{SUSD ObEx 05}. 
Thick continuous and dashed curves show the increase of the soliton amplitude in accordance with equation found above. 
The decrease of width of the soliton is obtained either. 
Our estimations show that the negative value of $\chi$ cannot be reached in the isotropic regime, 
but it can be obtained in the anisotropic case.

Let us assume that $c_{0}=0$.
Hence, parameter $\mid E\mid=v_{0}^{2}/2$ is also simplified.
Parameters $\alpha$ and $\chi$ transform into $\alpha_{0}\equiv\alpha(c_0=0)=mv_0\eta/\hbar$
and $\chi_0\equiv\chi(c_0=0)=16m^{2}v_0^{2}g_{2}/\hbar^{2}g$.
Parameter $\chi_0$ is proportional to higher degree of the soliton velocity $v_{0}^{2}$.
Hence, the relative contribution of the TOIR effects (which are proportional to $\chi_0$) grows at the increase of the soliton velocity.
The amplitude $n_{max}$ grows either.

\subsection{Anisotropic case}

General structure of equations (\ref{KdV eq in the third order appr form})-(\ref{KdV eq in the first order sol}) conserves for the nonisotropic case,
but it has a different meaning.
Changing are coming via different and anisotropic behavior of parameter $g_{2}=g_{2}(\theta)$.

Parameter $g_{2}$ enters equation (\ref{KdV eq in the first order sol}) via $\chi$ only.
Trace the charge of the soliton profile as a function of $\theta$.

If $G_{2,2}$ is relatively small, as in Fig. \ref{SUSD ObEx 03}, 
parameter $G_{2}$ has small change as the function of the angle $\theta$. 
the Fig. \ref{SUSD ObEx 03} demonstrates the twofold decrease of the $G_{2}$ at change of angle $\theta$ from $0$ to $\pi/2$. 
Moreover, the sign of $G_{2}$ does not change. 

The regime of large $G_{2,2}$ is demonstrated in Fig. \ref{SUSD ObEx 04}. 
The parameter $G_{2}$ changes from $0$ to $4$. 
Hence, it changes on several order on magnitude. 
Furthermore, the sign of $G_{0}$ changes in this regime. 
At $G_{2}<0$ the magnitude of $\mid G_{2}\mid$ reaches $\mid G_{2}\mid=2$. 
It corresponds to the relatively large negative values of $\chi$ (an example is presented by the dashed curve in Fig. \ref{SUSD ObEx 05}).

\section{Bright soliton in repulsive BEC}

The section VII addresses the large amplitude solitons.
However, the nonperturbative methods are relatively complex.
Hence, we consider a method of the nonlinear analysis of the small amplitude perturbations.
There are several methods, but we use the reductive perturbation method \cite{Leblond JPB 08}, \cite{Washimi PRL 66}.
It is applied
for the analytical demonstration of the bright soliton existence in the repulsive 1D BEC.

The bright solitons in the attractive BEC described above is the well-known phenomenon 
while the repulsive BEC provides condition for the dark soliton.
Although, it is found that mediums with repulsive interaction (including repulsive BEC \cite{Andreev MPL B 12}) 
can support the bright solitons \cite{Andreev MPL B 12}, \cite{Wang NJP 14} 
under the influence of additional effects such as the quintic nonlinearity or the TOIR nonlocal nonlinearity. 
This phenomenon is studied here in terms of the developed model.

The reductive perturbation method introduces a dimensionless parameter $\varepsilon$.
It is an indicator of small value of the corresponding terms
(see below equations (\ref{TOIR Sol expansion 1D of conc}) and (\ref{TOIR Sol expansion 1D of vel})).
This parameter is also used for the scaling of space and time variables.
In our case the stretched variables include the expansion parameter in follows combination:
\begin{equation}\label{TOIR masht of var 1D}
\begin{array}{ccc}\xi=\varepsilon^{1/2}(h-ut),&\tau=\varepsilon^{3/2}ut, &\end{array}\end{equation}
where $u$ is the phase velocity of the wave,
$\varepsilon$ is a small nondimension parameter.

An operational relations are arisen from (\ref{TOIR masht of var
1D})
\begin{equation}\label{TOIR repres. of var. 1D}
\begin{array}{ccc} \partial_{h}=\varepsilon^{1/2}\partial_{\xi},
&\partial_{t}=u (\varepsilon^{3/2}\partial_{\tau}-\varepsilon^{1/2}\partial_{\xi} ). &\end{array}\end{equation}

The decomposition of the concentration and velocity field involves
a small parameter $\varepsilon$ in the following form:
\begin{equation} \label{TOIR Sol expansion 1D of conc} n=n_{0}+\varepsilon n_{1}
+\varepsilon^{2}n_{2}+... ,\end{equation}
and
\begin{equation} \label{TOIR Sol expansion 1D of vel} v_{h}=\varepsilon v_{1}
+\varepsilon^{2}v_{2}+... .\end{equation}
Presented in (\ref{TOIR
Sol expansion 1D of conc}) equilibrium concentration $n_{0}$ is a
constant. We put   expansions (\ref{TOIR repres. of var.
1D})-(\ref{TOIR Sol expansion 1D of vel}) in equations (\ref{continuity h}) and (\ref{Euler h}).
Consequently, the system of equation is divided
into systems of equations in different orders on $\varepsilon$.

Equations emerging in the first order by $\varepsilon$ from the
system of equations (\ref{continuity h}) and (\ref{Euler h}) have the following form
\begin{equation} \label{} u\partial_{\xi}n_{1}-n_{0}\partial_{\xi}v_{1}=0,\end{equation}
and
\begin{equation} \label{TOIR Sol 1 order syst} mun_{0}\partial_{\xi}v_{1}=g n_{0}\partial_{\xi}n_{1}, \end{equation}
and lead to the following expression for the phase velocity $u$:
\begin{equation} \label{TOIR Sol rel for U 1D} u^{2}=\frac{g n_{0}}{m}.\end{equation}
Square of phase velocity $u^{2}$ should be positive.
Consequently $g$ is positive, i.e.
\begin{equation} \label{TOIR Sol cond of ex 1D} g>0.\end{equation}
It corresponds to the repulsive SRI.

We also obtain from (\ref{TOIR Sol 1 order syst}) a relation
between $n_{1}$ and $v_{1}$ and their derivatives
\begin{equation}\label{}\partial_{\xi}n_{1}=\frac{n_{0}}{u}\partial_{\xi}v_{1}.\end{equation}
Integrating this equation and using a boundary conditions
\begin{equation}\label{TOIR boundari cond the first} \begin{array}{cccc} n_{1} ,& v_{1}\rightarrow 0 & at &
x\rightarrow\pm\infty\end{array}\end{equation}
we have
\begin{equation}\label{TOIR 1e connection of var} n_{1}=\frac{n_{0}}{u}v_{1}.\end{equation}

In the second order by $\varepsilon$, from equations (\ref{continuity h}) and (\ref{Euler h}), we derive
\begin{equation}\label{TOIR 2e cont eq} -u\partial_{\xi}n_{2}+u\partial_{\tau}n_{1}+\partial_{\xi}(n_{0}v_{2}+n_{1}v_{1})=0,\end{equation}
and
$$-mu(n_{0}\partial_{\xi}v_{2}+n_{1}\partial_{\xi}v_{1}) +mun_{0}\partial_{\tau}v_{1}+mn_{0}v_{1}\partial_{\xi}v_{1}$$
\begin{equation} \label{TOIR Sol 2 order syst} -\frac{\hbar^{2}}{4m}\partial_{\xi}^{3}n_{1}= -g n_{0}\partial_{\xi}n_{2} -g n_{1}\partial_{\xi}n_{1} -g_{2}n_{0}\partial_{\xi}^{3}n_{1}.\end{equation}

We can express $n_{2}$ via $v_{2}$ and $n_{1}$, $v_{1}$ using
equation (\ref{TOIR 2e cont eq}) \textit{and} put it in equation
(\ref{TOIR Sol 2 order syst}). Using (\ref{TOIR Sol rel for U
1D}), we exclude $v_{2}$ from the obtained equation (\ref{TOIR Sol
2 order syst}). Thus, we obtain an equation which contains $n_{1}$
and $v_{1}$, only. Using (\ref{TOIR 1e connection of var}) and
expressing $v_{1}$ via $n_{1}$ we get the Korteweg--de Vries
equation for $n_{1}$
\begin{equation} \label{TOIR Sol KdV eq sos 1D}
\partial_{\tau}n_{1} +p_{1D}n_{1}\partial_{\xi}n_{1}+q_{1D}\partial_{\xi}^{3}n_{1}=0.\end{equation}
In this equation the coefficients $p_{1D}$ and $q_{1D}$ arise in
the form
\begin{equation} \label{TOIR Sol KdV koef 1D 1} p_{1D}=\frac{3}{2n_{0}},\end{equation}
and
\begin{equation} \label{TOIR Sol KdV koef 1D 2} q_{1D}=\frac{2m n_{0}g_{2}-\hbar^{2}}{4m n_{0}g}.\end{equation}

From equation (\ref{TOIR Sol KdV eq sos 1D})
we can find the solution in the form of a solitary wave using transformation $\kappa=\xi-V\tau$
\emph{and} taking into account boundary condition
$n_{1}=0$ and $\partial_{\kappa}^{2}n_{1}=0$ at
$\kappa\rightarrow\pm\infty$. 
As the result we find the bright soliton perturbation:
\begin{equation} \label{TOIR Sol solution of KdV 1D} n_{1}=\frac{2n_{0}V}{\cosh^{2}\biggl(\frac{1}{2}\sqrt{\frac{V}{q_{1D}}}\kappa\biggr)},\end{equation}
where $V$ is the velocity of soliton propagation to the right.
From expression $p_{1D}=3/2n_{0}$ and solution (\ref{TOIR Sol solution of KdV 1D})
we can find that a perturbation of concentration is positive.
Consequently, obtained solution is a bright soliton (BS).
A width of the soliton is given by equation $d=2\sqrt{q_{1D}/V}$.
The BS exists in the case $q_{1D}$ is positive.
From conditions $q_{1D}>0$ and (\ref{TOIR Sol cond of ex 1D})
we have
\begin{equation} \label{TOIR Sol solution existence cond} 2mn_{0}g_{2}-\hbar^{2}>0.\end{equation}
Relation (\ref{TOIR Sol solution existence cond}) is fulfil only in the case when $g_{2}$ is positive.
In the absence of the second interaction constant $g_{2}$ (i. e. in the
Gross--Pitaevskii approximation) the relation (\ref{TOIR Sol
solution existence cond}) does not fulfil and, consequently, the BS does not exist.
From equation (\ref{TOIR Sol solution existence cond})
we find that the second interaction constant $g_{2}$
should be positive and its module should be more than $\hbar^{2}/2mn_{0}$:
\begin{equation} \label{TOIR Sol solution existence cond sv} g_{2}>\frac{\hbar^{2}}{2mn_{0}}.\end{equation}

Relation (\ref{TOIR Sol solution existence cond sv}) corresponds to two times larger value of $g_{2,0}$ 
then the critical value of $g_{2,0}$ leading to domination of the TOIR over the quantum Bohm potential in the Bogoliubov spectrum (\ref{spectrum}) 
(for the isotropic case).

Condition (\ref{TOIR Sol solution existence cond sv}) can be considered explicitly for the anisotropic regime.
Use parameters $G_{2,0}$ and $G_{2,2}$ introduced in the end of Sec. V.
Then, the equation (\ref{TOIR Sol solution existence cond sv}) reappears as follows
\begin{equation} \label{anisotropic condition}
G_{2,0}-\frac{1}{\sqrt{5}}G_{2,2} \biggl[\sin^{4}\theta -\frac{1}{3}\sin^{2}\theta\cos^{2}\theta -2\cos^{4}\theta\biggr]>2 .
\end{equation}

Figs. \ref{SUSD ObEx 01} and \ref{SUSD ObEx 02} show that this condition gives a strong restriction on the direction of the soliton propagation.
Moreover, the existence of soliton itself can be restricted.

Fig. \ref{SUSD ObEx 02} shows that the propagation is possible in the small cone near the anisotropy axis.
While Fig. \ref{SUSD ObEx 01} demonstrates that in the chosen parameter regime the soliton existence is prohibited. 
Wherein $G_{2}$ is positive in Fig. \ref{SUSD ObEx 03} for all angles, 
but it is too small to fulfil condition (\ref{anisotropic condition}).

This is generalization of the result obtained earlier for the isotropic case \cite{Andreev MPL B 12}.
Moreover, the generalization of \cite{Andreev MPL B 12} for the boson-fermion mixtures is presented in Ref. \cite{Zezyulin EPJ D 13}.
Expectedly the boson-fermion mixture has more complex properties.
It demonstrates existence of the second kind solution found in Ref. \cite{Zezyulin EPJ D 13}
which has no analog in the single boson species.

\section{Conclusion}

A hydrodynamic model for the BEC with anisotropic short range interaction has been developed.

Bright solitons in the repulsive BEC are described.
They exist as a result of the SRI in the third order by the interaction radius.
Influence of the anisotropic SRI has been considered for this phenomenon.

Weakly anisotropic behavior of the traditional bright solitons in the attractive BEC has also been demonstrated.
The QHD method has been described in the paper allows to study any form of anisotropy of the SRI.
Being restricted by the TOIR, the dipole anisotropy is the single contribution which has been explicitly modeled in this paper.

The TIOR approximation reveals in the nonlocal form of the force field.
Hence, the force field is not just a gradient of function like for the GP approximation (the FOIR),
but it is the gradient of the Laplassian of a function.
Therefore, it contains higher derivatives of the concentration and it is more sensitive for the small scale perturbations.
Found form of the force field does not allow to derive the Cauchy-Lagrangian integral even for the isotropic SRI.
Consequently, an analog of the GP equation (a NLSE) does not exist either.
It shows the complexity of the model.
Anyway, the basic methods allow to study the fundamental properties of the system:
Bogoliubov spectrum of bulk collective excitations and possible solitons.

The anisotropy of the SRI requires a complex form of the electron subshells of valence electrons.
Hence, elements with incomplete $p$-, $d$-, $f$- subshells can show this property.
The modeling of lanthanides has demonstrated the contribution of anisotropic parts of the SRI in measurable properties.
Therefore, the presented hydrodynamic model is considered in context of these results.

The magnetic dipole-dipole interaction is cautiously neglected to stress our attention on the nonlocal and anisotropic parts of the meanfield SRI.
Modern models of BEC of lanthanides includes the quantum fluctuations allowing qualitative and quantitative modeling of the quantum Rosensweig instability.
The quantum fluctuations existing in the TOIR is the subject of future research
which allows a comprehensive analysis of instability of the cloud and regimes of stabilization.

Described in the paper approximate relations between the TOIR nonlocal nonlinearity and the quintic nonlinearity addressed to the understanding of the bright solitons in the repulsive BEC found by different groups and by different methods and in different physical systems.
Presented analysis provides a relation between whose models.

This relation gives a negative perspective for understanding of the quantum Rosensweig instability via the anisotropic SRI,
since it is now well-known
that the quintic nonlinearity does not give quantitative agreement with experiments
while the quantum fluctuations provide the required results.

Nevertheless, approximate analysis of the force field gives the following comparison between the quantum fluctuations, the TOIR, and quintic nonlinearity:
$n^{5/2}$, $\triangle n^{2}$, $n^{3}$ $\sim$ $n^{5/2}$, $ n^{8/3}$, $n^{3}$.

Therefore, the TOIR has an intermediate place in terms of the concentration dependence in compare with other two.
Moreover, the TOIR shows an different treatment of the different scales
since it contains derivatives in contrast with dependence on purely concentration.

Accordingly, the developed model opens a field of study 
which can accompany the dipolar BEC of lanthanides.

\section{Appendix A: General expressions for the QST up to the TOIR approximation}

\begin{widetext}
QST up to the third order on the interaction radius has the following form in terms of the microscopic many-particle wave function:
$$\sigma^{\alpha\beta}(\textbf{r},t)=-\frac{1}{2}\int dR\sum_{i,j.i\neq j}\delta(\textbf{r}-\textbf{R}_{ij})
r^{\beta}_{ij}\frac{\partial U(\textbf{r}_{ij})}{\partial r_{ij}^{\alpha}}\psi^{*}(R',t)\psi(R',t)$$
$$+\frac{1}{48}\partial_{\gamma}\partial_{\delta}\int dR\sum_{i,j,i\neq j}\delta(\textbf{r}-\textbf{R}_{ij}) r^{\beta}_{ij}r^{\gamma}_{ij}r^{\delta}_{ij}
\frac{\partial U(\textbf{r}_{ij})}{\partial r_{ij}^{\alpha}} \psi^{*}(R',t)\psi(R',t)$$
$$+\frac{1}{8}\int dR\sum_{i,j,i\neq j}\delta(\textbf{r}-\textbf{R}_{ij})\frac{r^{\alpha}_{ij}r^{\beta}_{ij}r^{\gamma}_{ij}r^{\delta}_{ij}}{r_{ij}}U(\textbf{r}_{ij})
\Biggl[[\partial_{\gamma1}\psi^{*}(R',t)]\partial_{\delta1}\psi(R',t)$$
$$+[\partial_{\gamma2}\psi^{*}(R',t)]\partial_{\delta2}\psi(R',t)-[\partial_{\gamma1}\psi^{*}(R',t)]\partial_{\delta2}\psi(R',t)
-[\partial_{\gamma2}\psi^{*}(R',t)]\partial_{\delta1}\psi(R',t)$$
$$+\frac{1}{2}\psi^{*}(R',t)\partial_{\gamma1}\partial_{\delta1}\psi(R',t)+\frac{1}{2}\psi^{*}(R',t)\partial_{\gamma2}\partial_{\delta2}\psi(R',t)-\psi^{*}(R',t)\partial_{\gamma1}\partial_{\delta2}\psi(R',t)$$
\begin{equation} \label{MKB sigma in 2 or} +\frac{1}{2}[\partial_{\gamma1}\partial_{\delta1}\psi^{*}(R',t)]\psi(R',t) +[\partial_{\gamma2}\partial_{\delta2}\psi^{*}(R',t)]\psi(R',t)-[\partial_{\gamma1}\partial_{\delta2}\psi^{*}(R',t)]\psi(R',t)\Biggr],\end{equation}
where $r_{ij}$ is the module $\textbf{r}_{ij}$,
$\partial_{\alpha1}$ and $\partial_{\alpha2}$ are the derivatives on $R_{ij}^{\alpha}$ located in the $i$-th and $j$-th places correspondingly.

The terms in the second order by the interaction radius are not demonstrated since their contribution is equal to zero
(in considering regime of the interaction anisotropy).
This conclusion appears at the integration over $\textbf{r}_{ij}$ (on the corresponding angle variables).


First two terms of equation (\ref{MKB sigma in 2 or}) contribute in the BEC state.
Other terms are related to the presence of particles in the excited states.
Let us illustrate it for the isotropic regime:
$$\sigma^{\alpha\beta}(\textbf{r},t)= \frac{1}{2}g\delta^{\alpha\beta}[2n_{BEC} n_{n} +2n_{n}^{2} +\sum_{g}n_{g}(n_{g}-1)|\varphi_{g}|^{4}]$$
$$+g_{2,0}\biggl[\frac{1}{6}(\delta^{\alpha\beta}\triangle+2\partial^{\alpha}\partial^{\beta}) [2n_{BEC} n_{n} +2n_{n}^{2} +\sum_{g}n_{g}(n_{g}-1)|\varphi_{g}|^{4}]$$
$$ -\frac{8}{\hbar^{2}}[mn_{BEC} (\delta^{\alpha\beta}\Pi_{n}^{\gamma\gamma} +2\Pi_{n}^{\alpha\beta} )
-\delta^{\alpha\beta} j_{BEC}^{\gamma}j_{n}^{\gamma} -2j_{BEC}^{\alpha} j_{n}^{\beta}]
+4n_{BEC} Tr[(\delta^{\alpha\beta}\partial^{\gamma}\partial_{\gamma}^{'}+2\partial^{\alpha}\partial^{\beta '})\rho_{n}(\textbf{r},\textbf{r}',t)]$$
$$-\frac{8}{\hbar^{2}}[mn_{n} (\delta^{\alpha\beta}\Pi_{BEC}^{\gamma\gamma} +2\Pi_{BEC}^{\alpha\beta})
-\delta^{\alpha\beta} j_{BEC}^{\gamma}j_{n}^{\gamma} -2j_{n}^{\alpha} j_{BEC}^{\beta}]
+4n_{n} Tr[(\delta^{\alpha\beta}\partial^{\gamma}\partial_{\gamma}^{'}+2\partial^{\alpha}\partial^{\beta '})\rho_{BEC}(\textbf{r},\textbf{r}',t)]$$
$$-\frac{8}{\hbar^{2}}[mn_{n} (\delta^{\alpha\beta}\Pi_{n}^{\gamma\gamma} +2\Pi_{n}^{\alpha\beta})
-\delta^{\alpha\beta} j_{n}^{2} -2j_{n}^{\alpha} j_{n}^{\beta}]
+4n_{n} Tr[(\delta^{\alpha\beta}\partial^{\gamma}\partial_{\gamma}^{'}+2\partial^{\alpha}\partial^{\beta '})\rho_{n}(\textbf{r},\textbf{r}',t)]$$
\begin{equation}\label{sigma in 2 or} +(\delta^{\alpha\beta}\delta^{\gamma\delta}+\delta^{\alpha\gamma}\delta^{\beta\delta}+\delta^{\alpha\delta}\delta^{\beta\gamma}) \sum_{g}n_{g}(n_{g}-1)[\varphi^{*}_{g} \varphi^{*}_{g} (\varphi_{g} \partial_{\gamma}\partial_{\delta}\varphi_{g} -\partial_{\gamma}\varphi_{g} \partial_{\delta}\varphi_{g})+c.c.]\biggr],\end{equation}
where
$\partial'_{\alpha}=\partial/\partial r'_{\alpha}$, and functions $n_{BEC}$, $n_{n}$, $j_{BEC}^{\alpha}$, $j_{n}^{\beta}$, $\Pi_{BEC}^{\alpha\beta}$, $\Pi_{n}^{\alpha\beta}$, $\varphi_{g}$ are functions of $\textbf{r}$ and $t$.
Subindex $BEC$ shows that the function describes the particles in the lower energy state.
Subindex $n$ shows functions related to the particles in the excited states.
Summation index $g$ is the full set of quantum numbers identifying the quantum states.
Functions $\varphi_{g}$are the microscopic wave functions of the single particle state with quantum numbers $g$.

The following representation of the hydrodynamic function is used in equation (\ref{sigma in 2 or}):
\begin{equation}\label{nvarphi} n(\textbf{r},t)=\sum_{g}n_{g}\varphi_{g}^{*}(\textbf{r},t)\varphi_{g}(\textbf{r},t),\end{equation}
\begin{equation}\label{jvarphi} j^{\alpha}(\textbf{r},t)=\frac{1}{2m}\sum_{g}n_{g}[\varphi_{g}^{*}(\textbf{r},t)\hat{p}^{\alpha}\varphi_{g}(\textbf{r},t)+\hat{p}^{\alpha *}\varphi_{g}^{*}(\textbf{r},t)\varphi_{g}(\textbf{r},t)] ,\end{equation}
\begin{equation}\label{Pivarphi}
\Pi^{\alpha\beta}(\textbf{r},t)=\frac{1}{4m}\sum_{g}n_{g}[\varphi_{g}^{*}(\textbf{r},t)\hat{p}^{\beta}\hat{p}^{\alpha}\varphi_{g}(\textbf{r},t)+(\hat{p}^{\beta
*}\varphi_{g}^{*}(\textbf{r},t))\hat{p}^{\alpha}\varphi_{g}(\textbf{r},t)+c.c.].
\end{equation}
\end{widetext}

Tensor $\Pi^{\alpha\beta}$ is the momentum flux.
Originally, $\Pi^{\alpha\beta}$ appears in the model at the derivation of the Euler equation (see eq. (5) in Ref. \cite{Andreev PRA08} or eq. (8) in Ref. \cite{Andreev PRB 11}).
Implicitly, tensor $\Pi^{\alpha\beta}$ can be found in equation (\ref{Euler v1}) as $mnv^{\alpha}v^{\beta}+p^{\alpha\beta}+T^{\alpha\beta}$,
while term $\partial_{\beta}(mnv^{\alpha}v^{\beta})$ is modified via the application of the continuity equation.

Function $2n_{BEC} n_{n} +2n_{n}^{2} +\sum_{g}n_{g}(n_{g}-1)|\varphi_{g}|^{4}$ presented in the first term in equation (\ref{sigma in 2 or}) comes from the calculation of $Tr(n_{2}(\textbf{r},\textbf{r}',t))$ in equation (\ref{sigma n_2+gen coef}).
Few intermediate steps of calculation are demonstrated in Sec. III of Ref. \cite{Andreev PRA08}.
The first term in equation (\ref{sigma in 2 or}) reproduces equation (30) of Ref. \cite{Andreev PRA08}.
It is mentioned in Ref. \cite{Andreev PRA08} that
$2n_{BEC} n_{n} +2n_{n}^{2}$ comes from the terms describing two interaction particles located in two different quantum states.
However, $2n_{n}^{2}$ requires contribution of pairs of particles in the same quantum states (with no contribution of the ground state).
Strictly, $n_{n}^{2}$ should be replaced by $n_{n}^{2}-\sum_{g,g\neq g_{0}}n_{g}^{2}|\varphi_{g}|^{4}$,
where $g_{0}$ corresponds to the ground state.
Hence, whole function transforms into
$2n_{BEC} n_{n} +2n_{n}^{2} -\sum_{g,g\neq g_{0}}n_{g}(n_{g}+1)|\varphi_{g}|^{4} + n_{g_{0}}(n_{g_{0}}-1)|\varphi_{g_{0}}|^{4}$.
The last term plays major role for the BEC
and it has the same form in both presentation.


\end{document}